\documentclass[12pt, oneside]{article}   	

\usepackage{jheppub}
\usepackage{amsmath}
\usepackage{amssymb}
\usepackage{amsfonts}
\usepackage{amsthm}
\usepackage{amsbsy}
\usepackage{verbatim}
\usepackage{array}
\usepackage{braket}
\usepackage[linesnumbered,lined,boxed,commentsnumbered]{algorithm2e}

\def\ben{\begin{equation}}
\def\een{\end{equation}}
\def\bea{\begin{eqnarray}}
\def\eea{\end{eqnarray}}

     \let\r=v

\def\ba{\begin{array}}
\def\ea{\end{array}}

\def\vx{\vec{x}}
\def\hlambda{\hat{\lambda}}
\def\hp{\hat{p}}

\def\bA{\bar{A}}
\def\trho{\tilde{\rho}}

\def\Xint#1{\mathchoice
   {\XXint\displaystyle\textstyle{#1}}%
   {\XXint\textstyle\scriptstyle{#1}}%
   {\XXint\scriptstyle\scriptscriptstyle{#1}}%
   {\XXint\scriptscriptstyle\scriptscriptstyle{#1}}%
   \!\int}
\def\XXint#1#2#3{{\setbox0=\hbox{$#1{#2#3}{\int}$}
     \vcenter{\hbox{$#2#3$}}\kern-.5\wd0}}

\def\dashint{\Xint-}

\def\Xsum#1{\mathchoice
   {\XXsum\displaystyle\textstyle{#1}}%
   {\XXsum\textstyle\scriptstyle{#1}}%
   {\XXsum\scriptstyle\scriptscriptstyle{#1}}%
   {\XXsum\scriptscriptstyle\scriptscriptstyle{#1}}%
   \!\sum}
\def\XXsum#1#2#3{{\setbox0=\hbox{$#1{#2#3}{\sum}$}
     \vcenter{\hbox{$#2#3$}}\kern-.4\wd0}}

\def\dashsum{\Xsum-}

\def\@fpheader{\ }

\title{\boldmath Finiteness of Entanglement Entropy in Collective Field Theory}

\author[a]{Sumit R. Das}
\author[b]{Antal Jevicki}
\author[b]{Junjie Zheng}

\affiliation[a]{Department of Physics and Astronomy, University of Kentucky, Lexington, KY 40506, USA}
\affiliation[b]{Department of Physics, Brown University, Providence, RI 02906, USA}

\emailAdd{sumit.das@uky.edu}
\emailAdd{antal\_jevicki@brown.edu}
\emailAdd{junjie\_zheng@brown.edu}

\abstract{We explore the question of finiteness of the entanglement entropy in gravitational theories whose emergent space is the target space of a holographic dual. In the well studied duality of two-dimensional non-critical string theory and $c=1$ matrix model, this question has been studied earlier using fermionic many-body theory in the space of eigenvalues. The entanglement entropy of a subregion of the eigenvalue space, which is the target space entanglement in the matrix model, is finite, with the scale being provided by the local Fermi momentum. The Fermi momentum is, however, a position-dependent string coupling, as is clear in the collective field theory formulation. This suggests that the finiteness is a non-perturbative effect. We provide evidence for this expectation by an explicit calculation in the collective field theory of matrix quantum mechanics with vanishing potential.
The leading term in the cumulant expansion of the entanglement entropy is calculated using exact eigenstates and eigenvalues of the collective Hamiltonian, yielding a finite result, in precise agreement with the fermion answer. Treating the theory perturbatively, we show that each term in the perturbation expansion is UV divergent. However the series can be resummed, yielding the exact finite result. Our results indicate that the finiteness of the entanglement entropy for higher dimensional string theories is non-perturbative as well, with the scale provided by the Newton constant.}
\dedicated{Dedicated to the memory of Ivan Andri\'
{c}.}
\begin{document} 
\maketitle
\flushbottom

\section{Introduction}

The entanglement entropy of a subregion in a relativistic quantum field theory is UV divergent because of short range correlations across the entangling surface. This is evident from the continuum limit of the earliest calculations on a lattice \cite{sorkin,srednicki}, or in calculations using the replica trick \cite{Callan:1994py, Holzhey:1994we,Calabrese:2009qy}. Another way to understand this divergence is to realize that the entanglement entropy of e.g. half space can be understood in terms of the standard thermodynamic entropy of quantum fields in Rindler space. The latter is divergent since the local temperature at the horizon diverges and the entropy of relativistic fields increases as a power of the temperature at high temperatures.

It is expected that if one can properly define entanglement entropy in string theory, the result should be finite \cite{Susskind:1994sm}. In string theory, it is not clear how one could go about defining the entanglement of a region in a precise manner. However, if one has access to a dual description in terms of a non-gravitational theory one could try to identify a quantity in the dual theory which provides a notion of entanglement in the gravitational theory in an appropriate approximation. 

This issue was addressed in \cite{Das:1995vj} for two-dimensional bosonic non-critical string theory, whose dual formulation is double-scaled gauged quantum mechanics of a single $N \times N$ Hermitian matrix $M$ with a Hamiltonian corresponding to an inverted harmonic oscillator \cite{c1review}. 
 As is well known, the singlet sector of the model becomes a theory of $N$ non-relativistic fermions in $1+1$ dimensions moving in this potential, whose coordinates are eigenvalues of the matrix. This can be in turn reexpressed as a second-quantized fermionic many-body theory, and in this case the notion of entanglement of a spatial region can be defined unambiguously. It was found in \cite{Das:1995vj} that when the external potential is absent, the leading term in the entanglement entropy for large enough interval $\Delta x$ is $\log (k_F \Delta x)/3$, where $k_F$ is the Fermi momentum. This is equal to the entanglement entropy of a {\em relativistic} massless scalar $1+1$ dimensions where the UV cutoff is replaced by the inverse Fermi momentum $k_F$. It was speculated in \cite{Das:1995vj} that in two-dimensional non-critical string theory, the UV cutoff will be the local Fermi momentum $k_F(x)$. A concrete calculation in the inverted harmonic potential was carried out in \cite{Hartnoll:2015fca} where it was indeed found that the cutoff is the position-dependent Fermi momentum.

From the point of view of the fermion theory, this is easy to understand. The behavior of the entanglement entropy $\sim \log [k_F (x) \Delta x]$ reflects the fact that the low energy excitations around the Fermi level have a linear dispersion relation, exactly like a massless relativistic boson with the speed of light replaced by the Fermi momentum. However, when the momenta becomes of the order of $k_F$ the quadratic term in the dispersion relation becomes important. Furthermore, the presence of a finite Fermi sea means that excitations have an effective UV cutoff given by the Fermi level. Likewise, at high temperatures the entropy of non-relativistic fermions increases logarithmically rather than a power law. In the Rindler calculation this leads to a finite contribution from the region near the horizon \cite{Das:1995vj}.

However, in the dual string theory $1/k_F (x)$ is proportional to the (position dependent) string coupling $g_{st}$. This means that the UV scale which makes the entanglement entropy finite is not simply the string length, but involves the string coupling. For the same reason, as emphasized in \cite{dastalk}, the finiteness of the entanglement entropy should be invisible in any finite order in perturbation theory \footnote{In \cite{dabholkar} the replica trick has been used to define an entanglement entropy in critical string theory. The string theory then lives on a cone, and the perturbative worldsheet partition function is finite with the string length providing the UV cutoff. While the relationship between this calculation and that of \cite{Das:1995vj,Hartnoll:2015fca} is not very clear, it appears that this calculation is quantifying a different kind of entanglement.}.

Generally collective field theory offers a reformulation of matrix and vector, providing a systematic $1/N$ expansion field theory. The two-dimensional string perturbation expansion likewise is generated through the collective field formulation. The dynamical degree of freedom in this case is a massless scalar field (``massless tachyon'') and one has an interacting quantum field theory. Although the higher string modes are non-propagating: they can, however, lead to non-trivial backgrounds. The collective field representation for the dynamics of the massless scalar is simply obtained by rewriting the matrix quantum mechanics
\ben
H = {\rm Tr} \left[ - \frac{1}{2} \left(\frac{\partial}{\partial M}\right)^2 +V(M) \right]
\label{0-4}
\een
in terms of the density of eigenvalues - the collective field $\phi(x)$ 
\ben
\phi (x) = {\rm Tr} \,\delta (x\cdot I - M) = \psi^\dagger (x) \psi (x)
\label{0-1}
\een
and its canonically conjugate momentum \cite{Jevicki:1979mb}.
This represents a non-relativistic bosonization and has been 
studied thoroughly.
Fluctuations of the collective field around its classical value behave as a massless scalar in $1+1$ dimensions with a position-dependent couplings proportional to inverse of the double-scaled Fermi level \cite{Das:1990kaa}. The space dimension descends from the space of eigenvalues. Thus to the lowest order in perturbation theory in this coupling the result for the entanglement entropy is UV divergent. The form of the answer from the fermionic description indicates that that this would continue to be divergent if one truncates perturbation theory to any finite order, as will be clear in the following. It is therefore natural to ask how does the finiteness of the entanglement entropy shows up in collective field theory. This is the central issue which we address in this paper \footnote{It turns out that the fluctuation of the collective field is related to the massless tachyon of string theory as defined in world-sheet string theory by a spatial transform (the leg pole transform) whose kernel is non-local at the string scale \cite{c1review}. This fact is not relevant to the discussion of whether the result is finite.}. 

It should be noted that the finiteness is due to the finiteness of the Fermi momentum $k_F$, which is proportional to the number density $N/L$ where $L$ is the size of a large box in which the fermions live. This means that the entanglement entropy remains finite in the limit $N \rightarrow \infty$, $L \rightarrow \infty$ with $N/L$ fixed. Likewise in the $c=1$ matrix model the quantity is finite in the double-scaling limit. What is important is that the coupling should be finite.

In this paper we calculate the entanglement entropy in the ground state {\em as defined in the fermionic many-body theory} using the collective field theory. For $d$-dimensional mutually non-interacting fermions, this entanglement entropy $S_A$ of a region $A$ has a well known expansion in terms of cumulants of the particle number \cite{klich},
\ben
S_A = {\rm lim}_{M \rightarrow \infty} \sum_{m=1}^{M} \alpha_{2m} (M) C_{2m},\quad C_m = (-i\partial_\lambda) \log \langle [ \exp(i\lambda N_A)] \rangle
\vert_{\lambda = 0} ,
\label{2-1a}
\een
where $N_A$ is the particle number operator in the region $A$, 
\ben
N_A = \int_A d^d\vx~ \psi^\dagger (\vx) \psi (\vx) .
\label{2-3}
\een
The coefficients $\alpha_{2m}$ are pure numbers given in \cite{klich}. In many situations $C_2$ is the leading contribution \footnote{There is no general proof of this: in fact there is no parametric suppression of the higher cumulants. However in many systems, including the systems considered in this paper the higher cumulants are nevertheless suppressed.}, and this is what we evaluate. Since the collective field is \eqref{0-1}, we have (using $\alpha_2 = \pi^2 / 3)$, for a single interval 
$a \leq x \leq b$ in one space dimension,
\ben
S^{(2)}_A = \frac{\pi^2}{3}\int_a^b dx \int_a^b dx^\prime \left[ \langle F| \phi(x) \phi(x^\prime) | F \rangle - \langle F| \phi(x) | F \rangle \langle F |\phi(x^\prime) | F \rangle \right] ,
\label{2-4}
\een
where $| F \rangle$ is the ground state. This is simply the integral of the connected Green's function. In this paper we calculate this quantity using the collective field theory Hamiltonian \footnote{For Slater determinant states all the terms in the cumulant expansion can be expressed in terms of the expectation value of the fermion phase space density \cite{dhl1,satya1}. This can be easily evaluated in a Thomas-Fermi approximation \cite{satya1,satya4,dhl2}. It remains to be seen if a theory of the phase space density regarded as an operator along the lines of \cite{dhar,son} is useful to proceed further.}. The quantity  $S^{(2)}_A$ is finite only if the short distance behavior of the collective field correlator is soft. 
As discussed above, in the lowest order in perturbation theory this correlator is exactly the same as that of a free massless relativistic scalar and therefore divergent.

Since the short distance behavior of the correlator is independent of the potential, we examine in detail the theory with no external potential. The interactions of the fluctuations of the collective field are then characterized by a coupling which is proportional to $1/k_F$.
In this case the exact eigenstates and eigenvalues of the collective Hamiltonian have been obtained in \cite{Jevicki:1991yi,nomura}. Using this exact solution we calculate the momentum space correlator and demonstrate agreement with the known answer obtained using fermions. We then calculate this quantity perturbatively and show that the perturbation expansion can be resummed. The resummed answer is in exact agreement with the result in the fermionic many-body theory which then leads to an agreement of the entanglement entropy. The expression (\ref{2-4}) involves an integral over the equal time correlator. We find that in momentum space the exact result is $|k|/\pi$ for $|k| < 2k_F$ which is also the leading perturbative result. For $|k| > 2k_F$ the result is a constant $2k_F / \pi$. The perturbation expansion in the collective field theory is a low momentum expansion in powers of $k/k_F$. The exact result shows that perturbatively there is no correction to the lowest order result which is independent of $k_F$. This means that the entanglement entropy is divergent perturbatively, and the finiteness of the result is a non-perturbative feature.

While our explicit calculation is for the matrix model without a potential, we expect that the same conclusion will hold in the presence of a potential, in particular the double-scaled $c=1$ matrix model. The collective field theory in these cases provides a field theory of strings with $1/N$ expansion being systematically generated \cite{Demeterfi:1991tz, Demeterfi:1991nw, Balthazar:2017mxh}. It needs to be treated with care and singular counter-terms present in the collective Hamiltonian will probably play a role.

As emphasized above, we are calculating the entanglement entropy as defined in the fermionic many-body theory, which we perform using collective field theory.
On the other hand one could define a notion of entanglement in the collective field theory itself. We need to determine if these two notions of entanglement agree with each other since bosonization involves a non-local transformation. This question has been investigated for lattice theories leading to relativistic fermions and conformal field theories in the literature \cite{ee-boson-fermion}, and is non-trivial when the subregion of interest consists of disconnected intervals. We will argue, however, that for non-relativistic fermions with a conserved fermion number, the situation is somewhat different. This is because now there is a first-quantized description, where the entanglement in the fermionic many-body theory becomes a target space entanglement \cite{target1}-\cite{target6}. In this first-quantized description, the operators which make sense are many-body operators involving a sum over all the identical particles. The latter can be in turn expressed either in terms of a second-quantized fermion field or in terms of the collective field and its momentum conjugate. For free fermions, and for Slater determinant states, it was shown in \cite{target2} that the reduced density matrix in the first-quantized description is exactly the same as that obtained in the second-quantized description. In the following we will argue that this implies that the entanglement entropy in terms of fermions is in fact the same as that in terms of the collective field.

While the singlet sector of single matrix quantum mechanics becomes a theory of free fermions, non-singlet sectors lead to models of interacting fermions, notably the Spin-Calogero models, particularly in the study of the long string sector \cite{Maldacena:2005hi,Balthazar:2018qdv}. Collective field theory for Calogero models have been developed in \cite{Andric:1982jk, Aniceto:2006rr,sen, Bardek:2010jg}. In these cases, the entanglement entropy can no longer be expressed in terms of fermion number cumulants. However, the collective formulation should be useful.

Our results should have implications for higher dimensional string theories whose holographic duals are matrix models with multiple matrices, e.g. the BFSS matrix model \cite{bfss} or the BMN matrix model \cite{bmn}. The notion of target space entanglement for multiple matrices has been formulated and explored in \cite{target2}-\cite{target5}. In terms of matrices explicitly, entanglement is discussed in \cite{target7,target8}. On the other hand, a collective formalism for the BMN matrix model has been established in \cite{bmncollective}. Here, the collective variables are ingredients of string fields. Since the gauge invariant matrix operators can be directly expressed in terms of the collective variables, a formulation of entanglement in terms of the latter will provide an understanding of the string theoretic meaning of target space entanglement.

In section (\ref{two}) we calculate the connected correlator of the collective field and hence the leading term in the entanglement entropy of a single interval for fermions without any external potential. The correlator is calculated first by using exact eigenstates and eigenvalues and then by resumming the perturbation expansion as well as exactly. In section (\ref{four}) we discuss the relationship of entanglement in the collective field and fermionic description. We also discuss possible applications to the long string sector which involves non-singlet states and multi-matrix models dual to higher dimensional strings. Section (\ref{five}) contains a discussion. The appendix provides some details of the derivation of the expression of the exact eigenstates and eigenvalues of the Hamiltonian.

\section{Entanglement Entropy for a Vanishing Potential}
\label{two}

In this section we consider the singlet sector of matrix quantum mechanics (\ref{0-4}) and the associated entanglement entropy. 
In the collective field formalism, the Hamiltonian is given by \footnote{In addition the general collective Hamiltonian contains a singular subleading counterterm. In this case this counterterm does not play much of a role, except to ensure that the $O(1/N^2)$ corrections to the ground state energy vanish.}
\ben
H = \frac{1}{2} \int dx ~\left[ \partial_x \Pi\,\phi\,\partial_x\Pi +\frac{\pi^2}{3}\phi^3 - 2\mu_F \phi \right]
\label{2-1}
\een
where $\Pi(x)$ is the canonically conjugate momentum to $\phi(x)$ defined in (\ref{0-1}) and $\mu_F$ is a Lagrange multiplier which imposes the condition 
\begin{equation}
    \int dx \,\phi(x) = N .
\end{equation}

The classical solution features a uniform distribution
\begin{equation}
    \phi_0=\frac{k_F}{\pi}, \quad\mu_F = \frac{1}{2}k_F^2.
\end{equation}
To study quantum fluctuations, we expand the collective field around the classical solution
\begin{equation}
    \phi(x)=\phi_0+\eta(x), \quad \partial_x\Pi(x)\rightarrow \partial_x\Pi(x).
\end{equation}
The fluctuation Hamiltonian becomes
\begin{equation}
    H=\frac{1}{2}\int dx \left\{\frac{k_F}{\pi}\left[(\partial_x\Pi)^2+(\pi\eta)^2\right]+\left[\partial_x\Pi\,\eta\,\partial_x\Pi+\frac{1}{3}(\pi\eta)^3\right]\right\}.
    \label{hamil}
\end{equation}
Writing $\eta = \partial_x \varphi$ and $\partial_x \Pi = \Pi_\varphi$ we see that it is evident that such a perturbation expansion is essentially a low energy expansion. For a process with momentum $k$ we see that (\ref{hamil}) is a theory of a massless scalar field in $1+1$ dimensions with cubic interactions. The cubic terms are small when the momenta $k$ are small compared to $k_F$ so that there is a perturbative expansion in powers of $k/k_F$
The quantity $S_A^{(2)}$ can be now expressed entirely in terms of the connected equal time
Green's function $G(x,x^\prime) \equiv \braket{\eta(x)\eta(x')}$ leading to
\ben
S_A^{(2)} = \frac{\pi^2}{3}\int_a^b dx \int_a^b dx' \,G(x,x') .
\label{2-7}
\een
In lowest order in the perturbation expansion, the Green's function is that of a massless field, so that the coincident integrated Green's functions which appear in (\ref{2-7}) are logarithmically divergent, leading to a logarithmically divergent result for  $S_A^{(2)}$ - exactly as expected.
The detailed form of $S_A^{(2)}$ depends on the boundary conditions. For example when the theory lives in a large box of size $L$, the integrated Green's function is
\begin{equation}
    \int dx \int dx'\, G(x,x')=-\frac{1}{2\pi^2}\log\left|x-x'\right|.
\end{equation}
leading to the entropy
\begin{equation}
S_A^{(2)}=\frac{1}{3}\log\frac{b-a}{\epsilon}.
\end{equation} 

On the other hand, the answer in the fermionic many-body theory is not divergent. The underlying reason is the fact that the fermions are non-relativistic.
Fluctuations of the collective field are particle-hole pair excitations around the Fermi sea. In the exact theory the energy of such an excitation is 
\ben
\omega = k_F \left( k + \frac{1}{2 k_F} k^2 \right) .
\label{2-8}
\een
The perturbative spectrum of the collective field is linear. As expected, this captures only the low energy spectrum, valid for $k \ll k_F$. On the other hand, the divergence of the entanglement entropy comes from the UV. Since $1/k_F$ is the coupling constant in the collective theory, this would mean that the correct answer with a finite $k_F$ has to be non-perturbative in the collective theory. 

In the next subsection we will demonstrate the exact spectrum with eigenstates in the collective formulation \cite{Jevicki:1991yi} and \cite{nomura}. The result is complete agreement with the known result in the fermionic many-body theory, featuring the poles corresponding to (\ref{2-8}). This means that at non-perturbative level the finite entropy is obtained in an exact calculation. We then consider the theory perturbatively. We show that the perturbation expansion can be resummed, again yielding the exact result.

\subsection{Direct evaluation using exact eigenstates}
\label{sec:2.1}

In this section we will obtain the Green's function of the collective field using exact eigenstates of the full Hamiltonian, using \cite{Jevicki:1991yi}. If we express the Hamiltonian as $H = H_2 + H_3$, where $H_2, H_3$ are the quadratic and cubic parts, it follows from the commutation relations of $\alpha_{L,R}$ that 
\begin{equation}
 [ H_2 , H_3 ] = 0   
\end{equation}
so that they can be simultaneously diagonalized. The eigenstates of $H_2$ are characterized by the total momentum $k$ in the emergent space direction $q$ which can be distributed among any number of particles in multiple ways. Thus these eigenstates are degenerate. It is useful to consider the coordinate $q$ to be in a periodic box of length $L$ so that the momenta 
\ben
k = \frac{2\pi n}{L},\quad n = 0, \pm 1, \pm 2, \cdots .
\label{3-10}
\een
Then the degeneracy of $H_2$ can be characterized by partitions of an integer. The exact eigenstates of $H$ are then obtained by transforming to a basis which also diagonalizes $H_3$.

The construction of exact eigenstates and eigenvalues follows from the connection of the matrix model Hamiltonian and the Laplacian on $U(N)$. Consider the unitary matrix $U$
\ben
U = {\rm exp} \left( \frac{2\pi i}{L} M \right) .
\label{3-1}
\een
Then the Hamiltonian is given by
\ben
H = -\frac{1}{2} {\rm Tr} \left( \frac{\partial}{\partial M}\frac{\partial}{\partial M} \right) = \left(\frac{2\pi}{L} \right)^2 \sum_\alpha C_\alpha C_\alpha ,
\label{3-2}
\een
where 
\ben
C_\alpha = {\rm Tr} \left( t^\alpha \frac{\partial}{\partial U} \right) .
\label{3-3}
\een
Here $t^\alpha, \alpha = 1 \cdots N^2$ are the generators of $U(N)$. 
The Hamiltonian is therefore the Laplacian on $U(N)$.
Let us introduce the collective variables
\begin{equation}
    \phi_n={\rm Tr}\,U^n.
\end{equation}

These are Fourier transforms of the collective field $\phi (x)$ of the previous section. Using the standard procedure in \cite{Jevicki:1979mb} the collective Hamiltonian is
\begin{equation}
\label{collh}
\begin{split}
    H_2=&\frac{2\pi(N-1)}{L}\sum_n |n|\phi_n\frac{\partial}{\partial \phi_n},\\
    H_3=&\frac{1}{2} \left(\frac{2\pi}{L}\right)^2
    \sum_{n,m} nm\phi_{n-m}\frac{\partial}{\partial \phi_n}\frac{\partial}{\partial \phi_m}+\sum_{n,m}|n|\phi_m\phi_{n-m}\frac{\partial}{\partial \phi_n}.
\end{split}
\end{equation}

The eigenstates can be now expressed in terms of characters of representations of $U(N)$. Consider a representation described by a Young tableau with $n$ boxes with $\lambda_j$ boxes in the $j$-th row
\ben
\lambda \equiv \{\lambda_1,\lambda_2,\cdots \},\quad\lambda_1 > \lambda_2 \geq \lambda_3 \geq \cdots, \quad \sum_j \lambda_j = n.
\label{3-11}
\een
The eigenstates of $H$ are then given by the Schur polynomials of $(\phi_1 \cdots \phi_n)$. Denote a conjugacy class of the permutation group $S_n$ by
\ben
\nu=\{1^{\nu_1}, 2^{\nu_2},\cdots\}.
\een
This corresponds to a partition of $n$ where $j$ appears $\nu_j$ times.
Then the Schur polynomials may be written as 
\begin{equation}
\label{egfn}
    s_\lambda(\{\phi\})=\sum_{\nu}\chi^\lambda_\nu \prod_{m}\frac{\phi_m^{\nu_m}}{\nu_m!m^{\nu_m}},
\end{equation}
where $\chi^\lambda_\nu$ denotes the character of the irreducible representation $\lambda$ for $\nu$ of $S_n$. This Fock space representation of this state may be obtained by the representation
\ben
\phi_n \rightarrow \sqrt{n}a^\dagger_n,\quad \frac{\partial}{\partial\phi_n}\rightarrow \frac{1}{\sqrt{n}}a_n,\quad[ a_m , a^\dagger_n ] = \delta_{mn}.
\een
and the fluctuation of the collective field is
\ben
\delta \phi_n = \int dx \,e^{-\frac{2\pi i n}{L}}\,\eta(x) = \sqrt{n} (a_n + a_n^\dagger) .
\label{3-15}
\een
In terms of these annihilation and creation operators the Hamiltonian reads
\begin{equation}
\begin{split}
    H_2=&\frac{2\pi}{L}k_F\sum_{n\neq 0}|n|a_n^\dagger a_n ,\\
    H_3=&\frac{2\pi^2}{L^2}\sum_{n,m>0;n,m<0}\sqrt{nm|n+m|}(a^\dagger_na^\dagger_ma_{n+m}+a^\dagger_{n+m}a_na_m) .
    \label{3-15a}
\end{split}
\end{equation}
The eigenstate in question is then expressed in terms of the Fock vacuum $|0 \rangle$
\begin{equation}
    \ket{\lambda}=s_\lambda(\sqrt{j}a^\dagger_j)\ket{0}.
\end{equation}
The eigenvalue of the Hamiltonian $H$ can be then computed to yield
\bea
E_{\lambda} & =&  E_2 + E_3 ,
\nonumber \\
E_2 & = & \frac{1}{2} \left(\frac{2\pi}{L} \right)^2 Nn,\quad E_3 =\frac{1}{2} \left(\frac{2\pi}{L} \right)^2 \sum_{j} \lambda_j (\lambda_j-2j+1).
\label{3-16}
\eea
In this equation $E_2$ is the eigenvalue of $H_2$ and $E_3$ is the eigenvalue of $H_3$.

A particular class of these states play a special role in the following. These are single particle states. For a given $n$ these states are labelled by an integer $m$, leading to a $\lambda$ given by
\ben
\lambda (n,m) =\{m+n-M,\underbrace{1,1,\cdots,1}_{M-m+1}\}.
\label{3-18}
\een


Using (\ref{3-16}) the energy of this state above the ground state is given by
\ben
E_\lambda(n,m) =\frac{1}{2}\left(\frac{2\pi}{L} \right)^2 (n^2+2nm) .
\label{3-17}
\een
In terms of continuous momenta $k = 2\pi n / L$, $p= 2\pi m / L$ for a large box, we have 
\ben
E_{\lambda (p,k)} = \frac{1}{2} (k^2 + 2 pk) = \frac{1}{2} [ (p+k)^2 - p^2 ] .
\label{3-20}
\een
Similarly for negative $k$ we have the particle-hole branch which has the dispersion relation (\ref{3-20}) with $k \rightarrow -k$.

The Weyl formula expresses Schur polynomials as ratios of Slater determinants - this means that these exact eigenstates are precisely states of an $N=2M+1$ fermionic many-body theory \cite{Jevicki:1991yi}. The ground state is the filled Fermi sea where the states labelled by $-M,-M+1,\cdots M$ are filled. The Fermi momentum $k_F$ is given by
\ben
k_F = \frac{\pi(N-1)}{L} = \frac{2\pi M}{L} .
\een
The state represented by (\ref{3-16}) is a state where the a fermion is removed from the $m$-th level and moved to the $(n+m)$-th level. Note that the collective Hamiltonian  (\ref{3-15a}) is the Hamiltonian of fluctuations so that the energies are the excitation energies of the fermionic many-body theory. This correspondence immediately implies that the states $|\lambda(p,k) \rangle$ are the only states which have non-vanishing matrix elements 
\ben
\langle 0 | \delta \phi (k) | \lambda \rangle .
\een
Without any reference to the fermions, this result can be proven as follows.
According to the Frobenius characteristic formula, in order to give single particle states, the cycle type must be
\begin{equation}
    \nu=\{n^1\}.
\end{equation}
Therefore, any Schur polynomial $s_\lambda$ with non-vanishing character $\chi^\lambda_\nu$ of the particular cyclic type $\nu$ contributes to the Dirac bracket. We can compute $\chi^\lambda_\nu$ using the Murnaghan–Nakayama rule
\begin{equation}
    \chi^\lambda_\nu=\sum_{Y\in \operatorname{BST}(\lambda,\nu)}(-)^{ht(Y)},
\end{equation}
where $\operatorname{BST}(\lambda,\nu)$ denotes all border-strip tableaux of the shape $\lambda$ and the type $\nu$, and $\operatorname{ht}(Y)$ denotes the sum of the heights of the border strips in $Y$. The height of a border strip is one less than the number of rows it touches. For a given Young tableau of the shape $\lambda$, we start to fill the boxes with $n$ integers $`1'$. Those Young tableaux not of the hook form must contain at least one $2\times 2$ square of $`1'$, thus they fail to form border-strip tableaux, which means the combination of $\lambda$ and $\nu$ gives 
\begin{equation}
    \chi^\lambda_\nu=0. 
\end{equation}
Hence only the Young tableaux of the hook form survive from the integral. In this case, the leading term of the Schur polynomial is equal to
\begin{equation}
    s_\lambda=(-)^{k_F-p}\frac{\phi_k}{k}+\cdots\rightarrow(-)^{k_F-p}\frac{1}{\sqrt{k}}a_k^\dagger+\cdots.
\end{equation}
We have identified $\phi_k$ with creation operator $\sqrt{k}a_k^\dagger$.

Consider now the two-point function of collective field fluctuations
\begin{equation}
        \tilde{G}(\omega,k)=\int dt \,e^{i\omega \tau}\braket{\rho(\tau,k)\rho(0,-k)}=\int_{-k_F}^{k_F}\frac{dp}{2\pi}\frac{|\bra{0}\delta \phi(k)\ket{\lambda(p,k)}|^2}{i\omega-E_\lambda(p,k)}.
\end{equation}
Using (\ref{3-20}), performing the integral, and adding the contributions for positive and negative $k$ we get
\begin{equation}
    \tilde{G}(\omega,k)=\frac{1}{2\pi k}\left(\log\frac{i\omega-k_F k+k^2/2}{i\omega-k_F k-k^2/2}-\log\frac{i\omega+k_F k+k^2/2}{i\omega+k_F k-k^2/2}\right).
    \label{8-1}
\end{equation}
After analytic continuation back to real time, this expression clearly displays the dispersion relation (\ref{2-8}) and is in exact agreement with a direct calculation in the fermionic many-body theory (see e.g. \cite{Pereira:2007}).

\subsection{Perturbative calculation and resummation}

\begin{figure}[h]
\centering
\includegraphics[width=8cm]{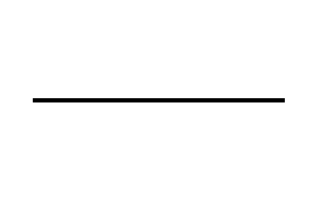}
\caption{Leading order}
\label{fig:f1}
\end{figure}

\begin{figure}[h]
\centering
\includegraphics[width=8cm]{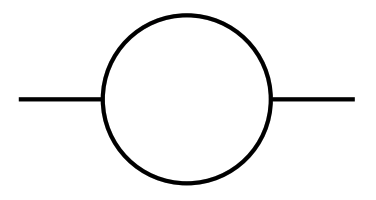}
\caption{Subleading order}
\label{fig:f2}
\end{figure}

\begin{figure}[h]
\centering
\includegraphics[width=8cm]{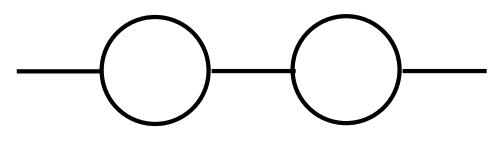}
\caption{Sub-subleading order:A}
\label{fig:f3}
\end{figure}

\begin{figure}[h]
\centering
\includegraphics[width=8cm]{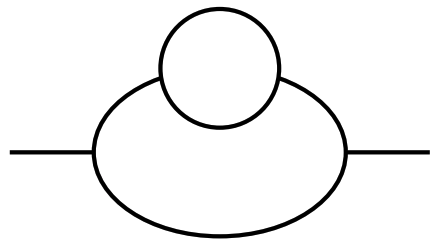}
\caption{Sub-subleading order:B}
\label{fig:f4}
\end{figure}

\begin{figure}[h]
\centering
\includegraphics[width=8cm]{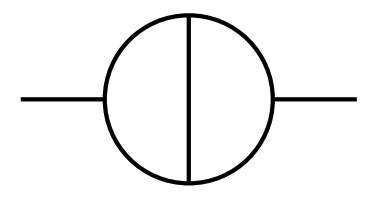}
\caption{Sub-subleading order:C}
\label{fig:f5}
\end{figure}

It is convenient to define left and right moving chiral bosons
\begin{equation}
    \alpha_L=\frac{1}{\sqrt{2\pi}}(\partial_x\Pi+\pi\eta), \quad \alpha_R=\frac{1}{\sqrt{2\pi}}(\partial_x\Pi-\pi\eta),
\end{equation}
with commutation relations
\bea
        \left[\alpha_L(x), \alpha_L(x')\right] &=& -i\partial_x\delta(x-x'),\\
        \left[\alpha_R(x), \alpha_R(x')\right] &=& +i\partial_x\delta(x-x'),\\
        \left[\alpha_L(x), \alpha_R(x')\right] &=& 0.
\eea
We can rewrite the Hamiltonian in terms of the new fields
\bea
    H&=& H_L+H_R,\\
    H_L&=&\frac{k_F}{2}\int dx \left(\alpha_L^2 + \frac{\sqrt{2\pi}g}{3k_F}\alpha_L^3\right),\\
    H_R&=&\frac{k_F}{2}\int dx \left(\alpha_R^2 - \frac{\sqrt{2\pi}g}{3k_F}\alpha_R^3\right).
\eea
Here we have introduced a small parameter $g$ to keep track of the terms in an perturbation expansion, which we will set to $1$ at the end of the calculation. As mentioned above the true expansion parameter is $k/k_F$ where $k$ is the momentum in the Green's function. The following calculation is similar to that in \cite{Pereira:2007}.
Using mode expansions
\begin{equation}
\alpha_{L,R}(\tau,x)=i\int_0^\infty dk\sqrt{\frac{k}{2\pi}}\left[a_{L,R}(k)e^{-k(k_F\tau\pm ix)}-a_{L,R}^\dagger(k)e^{k(k_F\tau\pm ix)}\right],
\end{equation}
we can compute the propagators of chiral bosons. In Euclidean signature,
\bea
        D_L(\tau,x)\equiv \braket{\alpha_L(\tau,x)\alpha_L(0,0)}&=&\frac{1}{2\pi}\frac{1}{(k_F \tau+ix)^2},\\
        D_R(\tau,x)\equiv \braket{\alpha_R(\tau,x)\alpha_R(0,0)}&=&\frac{1}{2\pi}\frac{1}{(k_F \tau-ix)^2}.
\eea
In momentum space, by doing contour integral we obtain
\begin{equation}
    D_{L,R}(\omega,k)=-\int_{-\infty}^\infty d\tau \int_{-\infty}^\infty dx \,e^{i(\omega \tau-kx)}D_{L,R}(\tau,x)=\frac{\mp k}{i\omega\pm k_F k}.
\end{equation}
Therefore we can read off the Feynman rules. Apart from propagators, the left and right vertices are given by $\pm\sqrt{2\pi}g$ respectively.
The main ingredient of calculating entanglement entropy is the Green's function of $\eta$, which we define in the following way
\begin{equation}
    \braket{\eta(\tau,x)\eta(0,0)}\equiv G(\tau,x)=-\int_{-\infty}^\infty\frac{d\omega}{2\pi}\int_0^\infty\frac{dk}{2\pi}\,e^{-i(\omega \tau-kx)}\tilde{G}(\omega,k),
\end{equation}
with
\begin{equation}
    \tilde{G}(\omega,k)=\frac{1}{2\pi}\left[\tilde{G}_L(\omega,k)+\tilde{G}_R(\omega,k)\right].
\end{equation}
The leading order $\{$\ref{fig:f1}$\}$ of $\tilde{G}_R(i\omega,k)$ is simply
\begin{equation}
   \tilde{G}_R^{(0)}(\omega,k)=D_R(\omega,k)=\frac{k}{i\omega-k_F k},
\end{equation}
while the subleading order $\{$\ref{fig:f2}$\}$ in $g$ gives
\begin{equation}
    \tilde{G}_R^{(1)}(\omega,k)=D_R(\omega,k)\Gamma_R(\omega,k)D_R(\omega,k).
\end{equation}
The self-energy can be computed again using contour integral
\begin{equation}
\begin{split}
    \Gamma_R(\omega,k)=&\frac{1}{2}\left(\sqrt{2\pi}g\right)^2\int_{-\infty}^\infty\frac{d\tilde{\omega}}{2\pi}\int_0^k \frac{d\tilde{k}}{2\pi}\,D_R(i\tilde{\omega},\tilde{k})D_R(i\omega-i\tilde{\omega},k-\tilde{k})\\
    =&\frac{g^2}{12}\frac{k^3}{i\omega-k_F k}.
\end{split}
\end{equation}
Plugging it back, we get
\begin{equation}
    \tilde{G}_R^{(1)}(\omega,k)=\frac{g^2}{24}\frac{k^5}{(i\omega-k_F k)^3}.
\end{equation}
The sub-subleading order $\{$\ref{fig:f3}, \ref{fig:f4}, \ref{fig:f5}$\}$ contains three Feynman diagrams, which give
\begin{equation}
 \frac{g^4}{144}\frac{k^9}{(i\omega-k_F k)^5}, \quad\frac{g^4}{504}\frac{k^9}{(i\omega-k_F k)^5}, \quad\frac{g^4}{280}\frac{k^9}{(i\omega-k_F k)^5}
\end{equation} 
respectively.


Collecting all of the contributions, we obtain
\begin{equation}
    \tilde{G}_R^{(2)}(\omega,k)=\frac{g^4}{80}\frac{k^9}{(i\omega-k_F k)^5}.
\end{equation}
This series can be resummed, leading to 
\begin{equation}
\begin{split}
    \tilde{G}_R(\omega,k)=&\tilde{G}_R^{(0)}(\omega,k)+\tilde{G}_R^{(1)}(\omega,k)+\tilde{G}_R^{(2)}(\omega,k)+\cdots\\
    =&\frac{1}{gk}\left[\frac{gk^2}{i\omega-kk_F}+\frac{1}{3}\left(\frac{gk^2}{i\omega-kk_F}\right)^3+\frac{1}{5}\left(\frac{gk^2}{i\omega-kk_F}\right)^5+\cdots\right]\\
    =&\frac{1}{gk}\log\frac{i\omega-k_Fk+g k^2/2}{i\omega-k_Fk-gk^2/2}.
\end{split}
\label{3-21}
\end{equation}
Sending $g$ to 1, we have
\begin{equation}
    \tilde{G}_R(\omega,k)=\frac{1}{k}\log\frac{i\omega-k_Fk+k^2/2}{i\omega-k_Fk-k^2/2}.
\end{equation}
Similarly, the Green's function of left chiral bosons is equal to
\begin{equation}
    \tilde{G}_L(\omega,k)=-\frac{1}{k}\log\frac{i\omega+k_Fk+k^2/2}{i\omega+k_Fk-k^2/2},
\end{equation}
thus
\begin{equation}
    \tilde{G}(\omega,k)=\frac{1}{2\pi k}\left(\log\frac{i\omega-k_Fk+k^2/2}{i\omega-k_Fk-k^2/2}-\log\frac{i\omega+k_Fk+k^2/2}{i\omega+k_Fk-k^2/2}\right).
    \label{2-11}
\end{equation}
This is in agreement with the exact answer (\ref{8-1}).

\subsection{Entanglement entropy of a single interval}

We are now ready to obtain an expression for the entanglement entropy of a single interval $(a,b)$, equation (\ref{2-7}). 
Notice that the only dependence on $x$ and $x^\prime$ is in the Fourier transformation, so we can integrate out them first,
\begin{equation}
    \int_a^b dx dx'\, e^{ik(x-x')}=\frac{4}{k^2}\sin^2\frac{k(b-a)}{2}.
\end{equation}
We now need to integrate (\ref{2-11}) to extract the equal time correlator. Performing a partial integration and using residue theorem, one gets
\begin{equation}
    \int_{-\infty}^\infty d\omega\log(\omega+ic)=2\pi|c| +I_1 +I_2 .
\end{equation}
Here 
\bea
I_1 & = &{\rm \lim}_{\Lambda \rightarrow \infty} [\omega \log(\omega + ic) - \omega]^\Lambda_{-\Lambda}, \nonumber \\
I_2 & = & ic\int_{\text{semicircle}} \frac{d\omega}{\omega + ic},
\eea
where the integral in $I_2$ is along an infinite radius semicircle in the lower- or upper-half plane depending on the sign of $c$.
In the integral over the four terms contained in $\tilde{G}(\omega,k)$ it may be checked that these divergent contributions cancel. 
After simplifying the expression, we obtain
\begin{equation}
\begin{split}
        S_A^{(2)}=&\frac{2}{3}\int_0^\infty \frac{dk}{k^3}\sin^2\frac{k(b-a)}{2}\left(\left|k_Fk+k^2/2\right|-\left|k_Fk-k^2/2\right|\right)\\
        =&\frac{2}{3}\int_0^{2k_F}\frac{dk}{k}\sin^2\frac{k(b-a)}{2}+\frac{4k_F}{3}\int^\infty_{2k_F}\frac{dk}{k^2}\sin^2\frac{k(b-a)}{2}.
\end{split}
\end{equation}
After performing the integral, the entanglement entropy can be recast into the form
\begin{equation}
\begin{split}
    S_A^{(2)}=&\frac{1}{3}\left\{-\operatorname{Ci}[2k_F(b-a)]-2k_F(b-a)\operatorname{Si}[2k_F(b-a)]+\log [k_F(b-a)]\right.\\
    &\left.+\pi k_F(b-a)+2\sin^2 [k_F(b-a)]+\gamma+\log 2\right\},
\end{split}
\label{4-1}
\end{equation}
where $\gamma$ is the Euler–Mascheroni constant. This is our final result. This, of course, is in exact agreement with the result obtained directly in the fermionic many-body theory.

It is now clear that this answer requires a resummation of the perturbative expansion. In fact, rather remarkably, the perturbative corrections to the leading order result exactly vanish. This may be seen explicitly by expanding the Fourier transform of the equal time correlator,
\ben
G_0(k) = \int_{-\infty}^\infty\frac{d\omega}{2\pi}\,\tilde{G}(\omega,k)=G^{(0)}_0(k)+G_0^{(1)}(k) + \cdots
\een
and using the expansion of $\tilde{G}(\omega,k)$. Performing the integrals explicitly one finds, for $k > 0$
\bea
G_0^{(0)} (k) & = & \frac{k}{\pi} , \nonumber \\
G_0^{(m)} (k) & = & 0 , \quad m=1,2,3 \cdots .
\label{3-22}
\eea
This means that there is no perturbative correction to the divergent lowest order result for the entanglement entropy. The answer is inherently non-perturbative. In fact the exact $G_0(k)$ obtained by integrating (\ref{2-11}) over $\omega$ is
\ben
G_0 (k) =
\left\{ \begin{array}{rcl}
|k|/\pi & \mbox{for}
& |k| < 2 k_F\\ 
2k_F/\pi & \mbox{for} & |k| > 2k_F 
\end{array}\right. .
\label{3-23}
\een
In position space, the exact equal time correlator is given by
\ben
G(x-y) = \left(\frac{k_F}{\pi}\right) \delta (x-y) \left( \frac{\sin [k_F(x-y)]}{k_F(x-y)}\right)-\left(\frac{k_F}{\pi}\right)^2 \left( \frac{\sin [k_F(x-y)]}{k_F(x-y)}\right)^2 .
\een
Integration of this quantity over the interval A yields the result (\ref{4-1}).

The perturbation expansion is in powers of $k/k_F$. Thus for all $k < 2k_F$ the result is indeed given exactly by the lowest order result (\ref{3-22}), consistent with what we found. The result for $k > 2k_F$, which is responsible for the finiteness of the entanglement entropy, is inaccessible in perturbation theory.

In the large interval limit $k_F(b-a)\gg1$, the entanglement entropy is given by
\begin{equation}
    S_A^{(2)}=\frac{1}{3}\{\log [k_F(b-a)]+1+\gamma+\log 2+\cdots\}.
\end{equation}
Notice that this result agrees with the lowest order calculation, except that the UV cutoff $\epsilon$ has been replaced by a finite number $1/k_F$. This can be understood as follows. In the large interval limit, the small momentum $G_0(k)$ contributes which can be calculated perturbatively. However the exact result (\ref{3-22}) shows that the low momentum behavior changes at $k \sim k_F$ - thus $k_F$ acts as a cutoff.

In the small interval limit $k_F(b-a)\ll1$, the entanglement entropy is given by
\begin{equation}
    S_A^{(2)}=\frac{1}{3}\{\pi k_F(b-a)+k_F^2(b-a)^2+\cdots\}.
\end{equation}
Unlike the large interval limit, this is extensive in the interval size. In this limit $G_0(k)$ is a constant so that the position space equal time correlator is a Dirac delta function, and the expression for the leading entanglement entropy (\ref{2-4}) leads to this extensive behavior.

\section{Entanglement in the Collective and Eigenvalue Descriptions}
\label{four}
As emphasized in the introduction, the preceding calculations are those of entanglement entropies as defined in the fermionic description, but calculated using the collective theory. In this section we discuss the connection of this quantity with the entanglement directly defined in terms of the bosonic collective field. This is the question of the relationship of the notion of entanglement of a region and bosonization.

In bosonization a fermion field is related to the boson field by a transformation which is non-local in space. Therefore, {\em a priori} the notion of locality in terms of bosons could be generally quite different for the notion of locality in terms of fermions and may lead to different entanglement entropies. For relativistic fermions and spin models this issue has been discussed in the literature \cite{ee-boson-fermion}.

For a non-relativistic fermionic many-body theory which is considered in this paper, the situation is rather different. This is because there is conserved number of fermions and a first-quantized description where the fermion coordinates are the dynamical variables. In fact, this is the basic description which comes from matrix quantum mechanics. Second-quantized fermionic many-body theory and collective field theory are two different formulations of this many-body system. The operators in the first-quantized formalism are of the type
\ben
{\cal O}_{mn} = \sum_{i=1}^N \hlambda_i^m \hp_i^n ,
\label{6-1}
\een
where $\hp_i$ are the momenta conjugate to the position operators $\hlambda_i$, and various orderings of these operators.
In terms of the second-quantized fermion field $\psi (x)$ this is
\ben
{\cal O}_{mn} = \int dx~ \psi^\dagger x^m (-i\partial_x)^n \psi,
\label{6-2}
\een
while the expression in terms of the collective field should be obtained by making a change of variables from $\{ \lambda_i \}$ to $\phi(x)$,
\ben
\phi(x) = \frac{1}{N} \sum_{i=1}^N \delta (x - \hlambda_i)
\label{6-3}
\een
and using the chain rule \cite{Jevicki:1979mb}.

The notion of entanglement of a subregion $A$ is best understood in terms of a subalgebra of operators. In the fermionic many-body theory the set of operators are simply those which are made out of the fermion fields $\psi (x), \psi^\dagger (x)$, with $x \in A$ . In the first-quantized language, specifying the subalgebra requires a constraint on the target space. This is best done by defining a projection operator for each $i$ \cite{target3}
\ben
P_i = \int_A dy~\delta(y - \hlambda_i) .
\label{6-4}
\een
The subalgebra of operators are then obtained by replacing 
\ben
(\hlambda_i, \hp_i) \rightarrow (P_i \hlambda_i P_i, P_i \hp_i P_i) .
\een
This projection breaks up the Hilbert space into a direct sum of super-selection sectors characterized by the number of particles $k$ which are in the subregion $A$. This is most easily seen by computing the expectation value of many-body operators of the form $\mathcal{O}_{mn}$ in some state described by a properly anti-symmetrized wavefunction $\Psi (\lambda_1,\lambda_2,\cdots \lambda_N)$. Consider operators of the form 
\ben
{\cal O}_{m} = \sum_{i=1}^N \hlambda_i^m ,
\label{6-5}
\een
whose projected version is
\ben
{\cal O}^P_{m} = \sum_{i=1}^N P_i\hlambda_i^m .
\label{6-5a}
\een
It is straightforward to see that the expectation value of the projected operator is
\ben
\langle \Psi |{\cal O}^P_m | \Psi \rangle 
= \sum_{k=1}^N {N \choose k} \sum_{i=1}^k \int_A \prod_{a=1}^k d\lambda_a \int_{\bA} \prod_{\alpha=k+1}^N d\lambda_\alpha~\Psi^\star (\lambda_1 \cdots \lambda_N) \lambda_i^m (\lambda_1 \cdots \lambda_N) 
\label{6-7}
\een
where $\bA$ is the complement of the region $A$. This contains a sum over the sectors mentioned above. In each sector labelled by $k$ the result can be obtained from an (unnormalized) reduced density matrix $\trho_k$ which is an operator in the $k$-particle Hilbert space of particles living in the region $A$. The trace ${\rm tr} \,\trho_k$ is the probability of finding $k$ particles in $A$.
The reduced density matrix of the entire system is block-diagonal 
\ben
\rho = {\rm diag}(\trho_1, \trho_2, \cdots \trho_N)
\label{6-8}
\een
and normalized (since the sum of probabilities is unity) and the target space entanglement entropy is given by the von Neumann entropy of $\rho$.

In the second-quantized fermionic many-body theory the Hilbert space is a direct product as usual and the reduced density matrix is obtained simply by integrating out the fermion fields in $\bA$. This density matrix which evaluates fermion bilinears in this region is identical to $\rho$ defined in (\ref{6-8}) - as was explicitly shown for Slater determinant states in \cite{target2}.

Let us now come to collective field theory. The subalgebra of operators pertaining to a region $A$ is the subalgebra of operators formed by taking products of $\{ \phi(x), \Pi (x) \}$, with $x \in A$. The restriction to $A$ can be implemented again by a projector, i.e.
\ben
\phi^P(x) = \int_A dy~ \delta (y-x) \phi (y) ,
\label{6-9}
\een
and similarly for $\pi^P(x)$. Focusing on many-body operators of the form (\ref{6-5}) the subalgebra now consists of operators of the form
\ben
{\tilde{\cal O}}^P_m = \int_R dx\, \phi^P(x) x^m = \sum_{i=1}^N \int_A dy~y^m\delta(y-\hlambda_i) .
\label{6-10}
\een
Clearly the subalgebra of operators ${\tilde{\cal O}}^P_m$ is identical to the subalgebra of operators ${\cal O}^P_{m}$, as may be checked explicitly by computing expectation values in arbitrary states. Therefore the reduced density matrices which evaluates these are identical as well. 

The same projector can be used to obtain the projected versions of many-body operators which involve momenta $p_i$. In terms of the collective field these involve integrals over the collective field and powers of the conjugate momentum, and the discussion above can be generalized.

We now discuss further applications of the collective field approach to the problem of target space entanglement in several problems of direct interest to string theory.

\subsection{Field theory of long strings}

A single matrix quantum mechanics with inverted oscillator potential $V(M)=-M^2/2$ is defined by the Hamiltonian
\begin{equation}
    H={\rm Tr} \left[-\frac{1}{2}\left(\frac{\partial}{\partial M}\right)^2-\frac{1}{2}M^2\right].
\end{equation}
Here $M$ is a Hermitian $N\times N$ matrix, which can be polar-decomposed in the form
\begin{equation}
    M=\Omega^\dagger\Lambda\Omega
\end{equation}
for some matrix $\Omega\in SU(N)/\mathbb{H}$ with $\mathbb{H}$ being the stablizer, where $\Lambda=\operatorname{diag}(\lambda_1,\cdots,\lambda_N)$ is diagonal. The invariance of the theory under $SU(N)$ transformation implies that we we can rewrite the Hamiltonian as
\begin{equation}
    H=-\frac{1}{2}\left(\frac{1}{\Delta}\sum_{i}\frac{\partial^2}{\partial\lambda_i^2}\Delta+\sum_{i}\lambda_i^2\right)+\dashsum_{i,j}\frac{Q_{ij}^\mathcal{R}Q_{ji}^\mathcal{R}}{(\lambda_i-\lambda_j)^2},
\end{equation}
associated with wavefunction $\Psi(\Lambda,\Omega)$ invariant under $S_N\ltimes U(1)^N$ gauge redundancy, where
\begin{equation}
    \Delta=\prod_{i<j}(\lambda_i-\lambda_j)
\end{equation}
is the Vandermonde determinant, and $Q_{ij}^\mathcal{R}$ is the $ij$ generator of $SU(N)$ under the representation $\mathcal{R}$.

While the singlet sector of matrix quantum mechanics reduces to a theory of {\em non-interacting} fermions, non-singlet sectors lead to {\em interacting} fermions whose coordinates are again given by the eigenvalues. In particular the long string sector is described by the adjoint sector and becomes related to the spin-Calogero model \cite{Aniceto:2006rr}. The problem of target space entanglement in the many-body quantum mechanics of these particles can be formulated exactly as above. In fact there is a well known collective field theory formulation of the Calogero model using its bosonized current-algebra representation, so that this can be reformulated in terms of collective fields \cite{Andric:1982jk,sen,Bardek:2010jg,Aniceto:2006rr}
\begin{equation}
\begin{split}
    H=&\int dx\left[\frac{1}{2}\partial_x\Pi(x)\phi(x)\partial_x\Pi(x)+\frac{\pi^2}{6}\phi^3(x)-\frac{1}{2}x^2\phi(x)-\partial_x\Pi(x)\bar{\psi}(x)\partial_x\psi(x)\right]\\
    &-\dashint dx dy \,\bar{\psi}(y)\frac{\phi(x)}{(x-y)^2}\psi(x)-\dashint dx\,\bar{\psi}(x)\left[\partial_x\dashint dy\,\frac{\phi(y)}{x-y}\right]\psi(x)\\
    &+\frac{1}{2}\dashint dx dy\,\bar{\psi}(x)\bar{\psi}(y)\left[\frac{\psi(x)-\psi(y)}{x-y}\right]^2 ,
\end{split}
\end{equation}
where boson $\phi$ and fermion $\psi$ represent closed string and long string respectively.

Since this is a model of interacting fermions, there is no direct connection between counting statistics \cite{Smith:2020gfl} and entanglement entropy. Nevertheless the short distance behavior of collective field correlators determines the behavior of the entanglement entropy. Preliminary results suggest that this can be obtained in a manner similar to the case detailed in this paper.

\subsection{Multi-matrix models and higher dimensional strings}

In \cite{target1,target2} the notion of target space entanglement has been generalized to multi-matrix models, e.g. the BFSS or the BMN matrix models. 

A Kaluza-Klein expansion of the $\mathcal{N}=4$ super-Yang-Mills theory on $\mathbb{R}\times \mathbb{S}^3$ in terms of spherical harmonics on $\mathbb{S}^3$ leads to matrix model reduction. Keeping only the zero mode degrees of freedom, for the Higgs sector the Lagrangian reads
\begin{equation}
    L=\operatorname{Tr}\left\{\frac{1}{2}\sum_i\left(\dot{\Phi}_i^2-\frac{1}{R^2}\Phi_i^2\right)+\frac{1}{4}\dashsum_{i,j}[\Phi_i,\Phi_j]^2\right\},\quad i,j=1,\cdots,6.
\end{equation}
The holomorphic notation introduces $SU(3)$ triplet $Z_i=\Phi_i+i\Phi_{i+3}$ and their complex conjugates $\bar{Z}_i$. Restriction to $1/2$-BPS configurations corresponds to single trace operators involving only the chiral primary operators of the general form $\operatorname{Tr}Z^n$. This model is essentially a one-matrix model described in this work. More generally addressing two-matrix problem, we may consider the simplest case of the complex matrix model with
\begin{equation}
    Z=A+B^\dagger, \quad \bar{Z}=A^\dagger+B.
\end{equation}

A gauge invariant notion of target space entanglement can be formulated in the following way. In a theory of several Hermitian matrices $M^I, I = 1 \cdots K$ consider a function $f(M)$ which is itself a Hermitian matrix. Then define a projector
\ben
P^f_{ij} = \int_A dy~\left[ \delta (y \cdot I - f(M)) \right]_{ij} .
\label{7-1}
\een
A set of gauge invariant operators are of the form
\ben
{\cal {O}}^{I_1I_2\cdots} = {\rm Tr} \left[ M^{I_1} M^{I_2} \cdots \right] .
\label{7-2}
\een
The projector (\ref{7-1}) can be then used to define a subalgebra of operators
\ben
{\cal {O}}_f^{I_1I_2\cdots} = {\rm Tr} \left[ P^fM^{I_1} P^fM^{I_2}P^f \cdots \right] .
\label{7-3}
\een
There is a reduced density matrix which evaluates expectation values of operators belonging to this subalgebra, and an associated entanglement entropy. This is a completely gauge invariant specification of the subalgebra.
In a gauge in which $f(M)$ is diagonal, the operator $P^f$ projects out the eigenvalues of $f(M)$ which lie in some specified interval $A$. A simple example involves $f(M) = M^1$. Then the eigenvalues of $M^1$ which lie in the interval $A$ are retained. In a sector where $n$ of the eigenvalues lie in $A$, $P^f$ projects onto an $n \times n$ block of the other matrices $M^I, I\neq 1$ the operator projects out. In the BFSS or BMN model, we have a $K$ dimensional target space $x^1 \cdots x^K$ and the eigenvalues of the matrices represent the locations of D0 branes in this target space. The reduced density matrix then evaluates measurements made on D0 branes whose $x^1$ lies in an interval A and the projection amounts to integrating out the open strings joining branes which do not lie in this region. Target space entanglement provides a concrete notion of entanglement in the string field theory dual to these matrix models.

The BMN matrix model has a collective field formulation \cite{bmncollective}. One has the general collective loops of $W, X$ and $Y$
\begin{equation}
    {\rm Tr}\,(W^{n_1}X^{m_1}Y^{k_1}W^{n_2}X^{m_2}\cdots) .
\end{equation}
These invariant loops variables denoted collectively by $\phi_C$ constitute all the observables in the full string field theory, where $C$ stands for word index. The collective Hamiltonian can be expressed in terms of $\phi_C$ and its conjugate $\pi_C$. The emergent spacetime is again seen through collective density \cite{Donos:2005vm}. In a way analogous to our treatment of the one-matrix problem one should then be able to consider entanglement in terms of this collective field. Potential evaluation of entanglement entropy can be done through numerical methods introducing in \cite{Koch:2021yeb}.

\section{Discussion}
\label{five}

In this paper we explored the question of finiteness of entanglement entropy in theories whose spatial dimensions emerge out of matrix degrees of freedom. More specifically, we addressed the question concretely in collective field theory of matrix quantum mechanics which becomes equivalent to two-dimensional non-relativistic fermions in an external potential. When the external potential is a regulated inverted harmonic oscillator this collective field theory is a string field theory of non-critical strings and the perturbation expansion is a string loop expansion. In the fermionic description the entanglement entropy is manifestly finite for a finite particle number density. However the collective field theory fluctuations are described by a self-interacting relativistic massless scalar field whose coupling is proportional to $k/k_F$. Thus to the lowest order in a perturbation expansion the result has the usual logarithmic divergence. The question is to understand how the interactions render the answer finite. This question is independent of the nature of the external potential.

We have answered this question unequivocally for the case where the potential is vanishing. In this case, the collective theory can be solved exactly \cite{Jevicki:1991yi} and the exact eigenvalues of the Hamiltonian are known to reproduce the fermion dispersion relation. Here we verified that the connected two-point function exactly reproduces the fermion four-point function. Since the leading term in the entanglement entropy involves an integral of this correlator this also leads to the correct exact answer. If we treat the interaction perturbatively we show that the leading order divergence is not cured in any finite order of the perturbation expansion. However the series can be resummed (as noted in \cite{Pereira:2007}) yielding the exact answer. The finiteness of the entanglement entropy is therefore essentially non-perturbative \footnote{It should be noted that we have performed a canonical perturbation expansion using the Hamiltonian. The conjugate momenta is non-polynomial in the time derivative of the field, so that there are an infinite number of vertices. A perturbation expansion using the Lagrangian will be rather complicated. However, a careful calculation should display cancellations and establish agreement with the Hamiltonian perturbative expansion.}.

When the external potential is an inverted harmonic oscillator, this system is a description of string field theory of bosonic non-critical string and the perturbation expansion is the string loop expansion. In this case, the collective field theory is more subtle. In particular in Hamiltonian has additional singular counter-terms which are subleading in $1/N$, and these are essential for a detailed correspondence to string theory. Nevertheless, we expect that the same mechanism will work in this case, i.e. the divergence is present in all finite orders of perturbation theory and its cure is non-perturbative. In the string theory this means that the scale which renders the entanglement entropy finite involves the string coupling $g_{st}$.

The main reason why the entanglement entropy is finite in the theories we consider is that the dynamics of the target space of these matrix models is non-relativistic in nature. This drastically alters the short distance behavior of correlations.
We have not performed any explicit calculation for theories with multiple matrices which are relevant to higher dimensional strings. However in known examples, e.g. the BFSS or BMN models, the target space is again non-relativistic. As conjectured in \cite{target2,target3} one would expect a similar mechanism for these models. 

In the examples we investigated in this paper, the finiteness of the entanglement entropy persists in a double-scaling limit where $N \rightarrow \infty$ and some other parameter (e.g. the size of the box for fermions in a box with no potential, or the Fermi level measured from the top of an inverted harmonic oscillator potential) also tuned keeping the coupling fixed. The treatment in section \ref{sec:2.1} is valid for any finite $N$. However only $N$ of the $\phi_m$ are independent variables because of trace relations. Naively this fact, also called the ``stringy exclusion principle", did not play a role in the subsequent analysis where we took both $N \rightarrow \infty$ and $L \rightarrow \infty$ keeping $k_F$ fixed. This point demands further investigation. It will be interesting to see if there are similar limits in these higher dimensional models. 

Finally it will be interesting to investigate the connection of the origin of finiteness discussed in this paper to other recent discussions based on the types of von Neumann algebras 
\cite{liu}.

\section{Acknowledgements} S.R.D. would like to thank Cesar Gomez, Gautam Mandal and Sandip Trivedi for discussions and Instituto de Fisica Teorica, Madrid for hospitality during the completion of this manuscript.  The work of S.R.D. is supported by National Science Foundation grants NSF/PHY-1818878 and NSF/PHY-2111673 and by the Jack and Linda Gill Chair Professorship. The work of A.J. and J.Z. is supported by the U.S. Department of Energy under contract DE-SC0010010.

\section{Appendix}

In this appendix we will provide the proof of the exact dispersion relation (\ref{3-16}). This result follows from two fusion rules. The first is the Littlewood-Richardson rule, which states that the fusion of Schur polynomials is determined by the equation
\begin{equation}
    s_\lambda s_\mu=\sum_\nu f^\nu_{\lambda,\mu}s_\nu,
\end{equation}
with coefficients $f^\nu_{\lambda,\mu}$ equal to the number of the Littlewood–Richardson tableaux of skew shape $\nu/\lambda$ and of weight $\mu$.
The second is the fusion rule of characters of permutation group $S_n$
\begin{equation}
    \frac{C_\rho}{d_\lambda}\chi^\lambda_\rho \frac{C_\mu}{d_\lambda}\chi^\lambda_\mu=\sum_\nu g^\nu_{\rho,\mu}\frac{C_\nu}{d_\lambda}\chi^\lambda_\nu,
\end{equation}
where the depth $d_\lambda$ of a Young tableau $\lambda$ is the number of boxes that do not belong to the first row. And the number of different permutations in the conjugacy class $\nu$ is given by
\begin{equation}
    C_\nu=\frac{n!}{\prod_j \nu_j!j^\nu},
\end{equation}
where $n!$ is the total number of elements in the permutation group $S_n$. The idea is to choose
\begin{equation}
    \rho=\left(1^{n-2},2^1\right).
\end{equation}
With this choice, we have
\begin{equation}
    C_\rho=n(n-1)
\end{equation}
and 
\begin{equation}
    \frac{C_\rho}{d_\lambda}\chi^\lambda_\rho=\sum_j\lambda_j(\lambda_j-2j+1)
\end{equation}
which is exactly the eigenvalue $E_3(\lambda)$ of $H_3$. We then have the eigenequation representing a special case of the multiplication formula
\begin{equation}
    H_3 C_\mu\chi^\lambda_\mu=\sum_\nu g^\nu_{\rho,\mu}C_\nu\chi^\lambda_\nu.
\end{equation}
Working out the special structure constant $g^\nu_{\rho,\mu}$ one gets
\begin{equation}
    \sum_k k\sum_{l=1}^{k-1}C_{\nu,s}\chi^\lambda_{\nu,s}+\sum_{k<l}kl C_{\nu,j}\chi^\lambda_{\nu,j},
\end{equation}
where 's' denotes splitting of the conjugacy class $C_\nu$
\begin{equation}
    \phi_k\rightarrow\phi_l, \phi_{k-l},
\end{equation}
while 'j' denotes joining of the conjugacy class $C_\nu$
\begin{equation}
    \phi_k, \phi_l\rightarrow \phi_{k+l}.
\end{equation}

\end{document}